\documentclass[aps,prb,twocolumn,superscriptaddress,showpacs,floatfix]{revtex4}

\usepackage{graphicx}
\graphicspath{{figs/}}
\begin{document}
\title{Elastic and non-linear stiffness of graphene: a simple approach}
\author{Jin-Wu~Jiang}
	\affiliation{Department of Physics and Centre for Computational Science and Engineering,
 		     National University of Singapore, Singapore 117542, Republic of Singapore }
\author{Jian-Sheng~Wang}
	\affiliation{Department of Physics and Centre for Computational Science and Engineering,
     		     National University of Singapore, Singapore 117542, Republic of Singapore }
\author{Baowen~Li}
        \affiliation{Department of Physics and Centre for Computational Science and Engineering,
                     National University of Singapore, Singapore 117542, Republic of Singapore }
        \affiliation{NUS Graduate School for Integrative Sciences and Engineering,
                     Singapore 117456, Republic of Singapore}
\date{\today}
\begin{abstract}
The recent experiment [Science \textbf{321}, 385 (2008)] on the Young's modulus and third-order elastic stiffness of 
graphene are well explained in a very simple approach, where the graphene is described by a simplified system and the force constant for the non-linear interaction is estimated from the Tersoff-Brenner potential.
\end{abstract}

\pacs{62.25.-g, 62.23.Kn, 81.05.Uw} \maketitle

Graphene has received a large number of interests since its discovery.\cite{Novoselov1, Berger} Among others, a recent experiment has shown that the graphene has excellent mechnical properties with very large Young's modulus.\cite{Lee} This experiment also found obvious non-linear effect for the graphene in the large strain regime. Theoretically, the elastic properties in the graphene can be studied
in the continuum mechanics approach\cite{Reddy} or the \textit{ab initio} method.\cite{Liu} In Ref.~\onlinecite{Pugno2009}, Pugno derived the third-order elastic stiffness (TOES) by relating it to the coefficient of thermal expansion\cite{Pugno2006}. In this paper, we describe the graphene in a simplified system with very simple interaction potential. There is
no variable parameters in the potential, where the non-linear interaction can be deduced from the Tersoff-Brenner (TB)
potential.\cite{Tersoff, Brenner} The recent experiment for the Young's modulus and the TOES of the 
graphene are well explained by our model.

\begin{figure}
  \begin{center}
    \scalebox{0.6}[0.6]{\includegraphics[width=7cm]{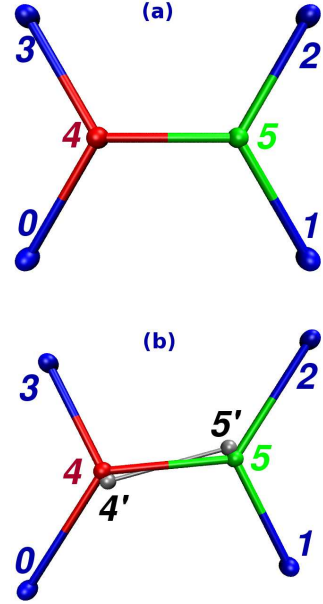}}
  \end{center}
  \caption{(Color online) Configuration for the simplified system. (a) is the equilibrium position. (b) is the configuration with strain. The inner atoms 4 and 5 are optimized into new equilibrium position $4'$ and $5'$ with four outside atoms fixed.}
  \label{fig_cfg}
\end{figure}

Graphene samples in the mechanical experiment\cite{Lee} usually have a radius larger than 0.75 $\mu m$. In this 
large radius two dimensional sheet, the number of atoms on the boundary ($N_{b}$) is much smaller than that of the inner atoms ($N_{i}$). It can be estimated as:
\begin{eqnarray*}
&N_{b}&=\frac{2\pi r}{b},\\
&N_{i}&=\pi r^{2}/(\frac{s}{2})=(\frac{r}{1.5\sqrt{3}b})(\frac{2\pi r}{b}),
\end{eqnarray*}
where $r$ is the radius of sample, and $b=1.42$~{\AA} is the C-C bond length in the graphene. $s=\frac{\sqrt{3}}{2}(\sqrt{3}b)^{2}$ is the area of the unit cell in the graphene. The ratio of these two numbers is:
\begin{eqnarray*}
&\frac{N_{b}}{N_{i}}&=1.5\sqrt{3}b/r\approx 5\times 10^{-4},
\end{eqnarray*}
where $r=0.75 \mu$m is used. It shows that the number of boundary atoms in the sample is about four orders lesser than the inner atoms. As a result, the contribution of the boundary atoms to the total energy is also four orders smaller than the inner atoms. In this sense, we can ignore the contribution of the boundary atoms to the total energy in the graphene. In the following we consider only the inner atoms.

Because of the translational symmetry, all unit cells in the graphene are equivalent to each other. We can consider only
one unit cell as a representative of the graphene, as shown in Fig.~\ref{fig_cfg}(a), where atoms 4 and 5 are the two non-equivalent carbon atoms in the representative unit cell.

In this paper, the direction with angle $\theta=0$ directs from atom 4 to 5 in Fig.~\ref{fig_cfg}(a). And $\theta=\pi/2$ is the vertical direction which is named armchair direction throughout this paper. 

\begin{table*}[t]
     \caption{Parameters in the Tersoff-Brenner potential. See text for the meaning of each parameters.}
     \label{tab_TB_par}
\begin{ruledtabular}
\begin{tabular}{|c|c|c|c|c|c|c|c|c|c|}
$D^{(e)}$ ($ev$) & S & $\beta$ (\AA$^{-1}$) & $R^{(e)}$ (\AA) & $R^{(1)}$ (\AA) & $R^{(2)}$ (\AA) & $\delta$ & $a_{0}$ & $c_{0}$ & $d_{0}$\\
\hline
6 & 1.22 & 2.1 & 1.39 & 1.7 & 2 & 0.5 & 0.00020813 & 330 & 3.5\\
\end{tabular}
\end{ruledtabular}
\end{table*}
The interaction potential we used includes both linear and non-linear terms. The linear interaction is the frequently used bond stretching interaction $V_{l}$:
\begin{eqnarray}
V_{l}=\frac{k_{l}}{2}(b-b_{0})^{2},
\end{eqnarray}
where, $b$ ($b_{0}$) is the strained (unstrained) C-C bond length in graphene. The force constant $k_{l}=305$Nm$^{-1}$ is taken from Ref.~\onlinecite{Jiang}, where this potential was applied successfully to explain phonon properties in the carbon nanotubes and graphene layers.

The non-linear interaction we applied has the form:
\begin{eqnarray}
V_{nl}=\frac{k_{nl}}{3}(b-b_{0})^{3},
\end{eqnarray}
and the constant $k_{nl}$ can be evaluated from the commonly used TB potential. Similar expansion has also been done in Refs.~\onlinecite{Huang, Mu}.

In the TB potential, the energy is expressed as:
\begin{eqnarray*}
V_{B}(r) & = & V_{R}(r)-\bar{B}_{ij}\cdot V_{A}(r).\end{eqnarray*}
$V_{R}$ and $V_{A}$ are the repulsive and attractive energy:
\begin{eqnarray*}
V_{R}(r) & = & \frac{D^{(e)}}{S-1}e^{-\sqrt{2S}\beta(r-R^{(e)})}f_{c}(r)\\
V_{A}(r) & = & \frac{D^{(e)}S}{S-1}e^{-\sqrt{2/S}\beta(r-R^{(e)})}f_{c}(r),\end{eqnarray*}
with the cut-off function $f_{c}(r)$:
\begin{eqnarray}
f_{c}(r) & = & \left\{
\begin{array}{cc}
1, &  \hspace{0.4cm}  r<R^{(1)}\; ,\\
\frac{1}{2}\left\{ 1+\cos\left[\frac{\pi(r-R^{(1)})}{R^{(2)}-R^{(1)}}\right]\right\}, &  \hspace{0.4cm} R^{(1)}<r<R^{(2)}\; ,\\
0, &  \hspace{0.4cm}  r>R^{(2)}\; .\\
\end{array}
\right.  \nonumber
\end{eqnarray}
The many-body coupling parameter is:
\begin{eqnarray*}
\bar{B}_{ij} & = & \frac{1}{2}(B_{ij}+B_{ji})\\
B_{ij} & = & \left[1+\sum_{k\not=ij}G(\theta_{ijk})f_{c}(r_{ik})\right]^{-\delta}.\end{eqnarray*}
The angle function $G(\theta_{ijk})$ is:
\begin{eqnarray*}
G(\theta_{ijk}) & = & a_{0}\left[1+\frac{c_{0}^{2}}{d_{0}^{2}}-\frac{c_{0}^{2}}{d_{0}^{2}+\left(1+\cos(\theta_{ijk})\right)^{2}}\right]\end{eqnarray*}
All parameters in the TB potential are listed in Table.~(\ref{tab_TB_par}). The ratio of $k_{nl}/k_{l}$ can be estimated from the TB potential as following.

(1). In the equilibrium structure of the graphene, all of the angles equal to $\frac{2}{3}\pi$. So the angle function $G(\theta_{ijk})$ can be simplified:
\begin{eqnarray*}
G(\theta) & = & a_{0}\left[1+\frac{c_{0}^{2}}{d_{0}^{2}}-\frac{c_{0}^{2}}{d_{0}^{2}+\left(1+\cos(\frac{2}{3}\pi)\right)^{2}}\right]\\
 & \approx & 0.037547\end{eqnarray*}

(2). Due to the small value of $G(\theta)$, we have:
\begin{eqnarray*}
B_{ij} & \approx & 1\\
\bar{B}_{ij} & \approx & 1.\end{eqnarray*}

(3). Because $R^{(2)}$=2~{\AA} in the cut-off function is about 40\% larger than the equilibrium bond length 1.42~{\AA} in the graphene, we simply set:
\begin{eqnarray*}
f_{c}(r) & \approx & 1.\end{eqnarray*}

(4). We expand the
exponent function in the repulsive and attractive energy in terms of $r-R^{(e)}$:
\begin{eqnarray*}
V_{R}(r) & = & \frac{D^{(e)}}{S-1}e^{-\sqrt{2S}\beta(r-R^{(e)})}\\
 & \approx & \frac{D^{(e)}}{S-1}\left[1-x+\frac{1}{2}x^{2}-\frac{1}{6}x^{3}\right]\\
\\V_{A}(r) & = & \frac{D^{(e)}S}{S-1}e^{-\sqrt{2/S}\beta(r-R^{(e)})}.\\
 & \approx & \frac{D^{(e)}}{S-1}\left[S-x+\frac{1}{2S}x^{2}-\frac{1}{6S^{2}}x^{3}\right],\end{eqnarray*}
with $x=\sqrt{2S}\beta(r-R^{(e)})$.

\begin{figure}
  \begin{center}
    \scalebox{1.2}[1.2]{\includegraphics[width=7cm]{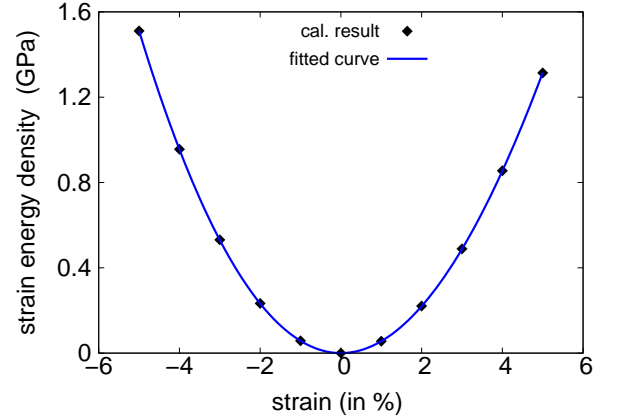}}
  \end{center}
  \caption{(Color online) The strain energy v.s strain. Strain energy at $\epsilon=-5\%$ is about 15$\%$ higher than that at $\epsilon=5\%$ implicating the non-linear interaction. The fitted function is $E(\epsilon)=\frac{1}{2}Y\epsilon^{2}+\frac{1}{3}D\epsilon^{3}$ with $Y=1.131$ TPa and $D=-2.360$ TPa.}
  \label{fig_energy}
\end{figure}
(5). Finally, the total binding energy is:
\begin{eqnarray*}
V_{B}(r) & \approx & V_{R}(r)-V_{A}(r)\\
 & \approx & \frac{D^{(e)}}{S-1}\left[(1-S)+\frac{S-1}{2S}x^{2}-\frac{1}{6}\frac{(S+1)(S-1)}{S^{2}}x^{3}\right]\\
 & \equiv & V_{0}+\frac{1}{2}k_{l}(r-R^{(e)})^{2}+\frac{1}{3}k_{nl}(r-R^{(e)})^{3}\end{eqnarray*}
where $V_{0}=D^{(e)}$. The vanish of the 1st order term in this final expression indicates that our above approximations are physically correct.

(6). Applying parameters in Table.~(\ref{tab_TB_par}), we obtain the value of $k_{nl}/k_{l}$:
\begin{eqnarray*}
\frac{k_{nl}}{k_{l}} & = & -\frac{1}{2}\times\frac{S+1}{S}\times\sqrt{2S}\beta\\
 & \approx & -3 \AA ^{-1}.\end{eqnarray*}
So, the non-linear constant is $k_{nl}=915$Nm$^{-1}\AA^{-1}$.

\begin{figure}
  \begin{center}
    \scalebox{1.2}[1.2]{\includegraphics[width=7cm]{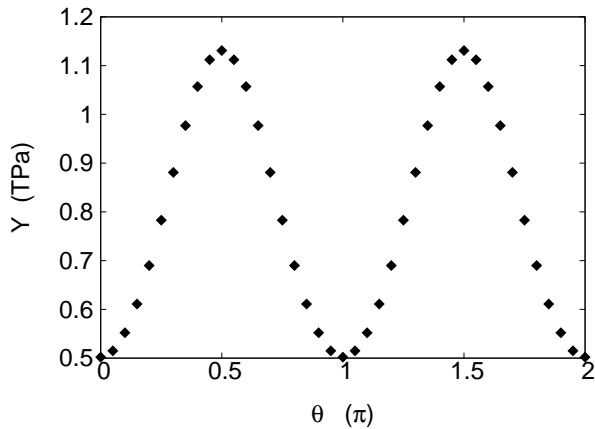}}
  \end{center}
  \caption{Young's modulus of the graphene.}
  \label{fig_Young}
\end{figure}
In the mechanical experiments on the graphene, the sample is stretched by force and the boundary is then fixed to measure the relation between the force and the strain. Similarly, as shown in Fig.~\ref{fig_cfg}, the system is strained (with strain $\epsilon$) in a particular direction (Fig.~\ref{fig_cfg}~(b)). Then the outer four atoms 0, 1, 2, and 3 are fixed, which simulates the experimental conditions. The two inner atoms 4 and 5 can move freely to achieve a new configuration $4'$ and $5'$ with minimum energy $E(\epsilon)$ . This optimized configuration is the equilibrium position for the system under strain.

Fig.~\ref{fig_energy} shows the relation between the density of strain energy and the strain. The strain in this figure is 
added along $\theta=\pi/2$ direction with the value in $[-5\%, 5\%]$, 
which is also the magnitude of strain added in the experiment.\cite{Lee} The non-linear effect can be clearly seen from this figure, since $E(-5\%)$ is about 15$\%$ higher than $E(5\%)$. Fitting this curve by function $E(\epsilon)=\frac{1}{2}Y\epsilon^{2}+\frac{1}{3}D\epsilon^{3}$, we can get the Young's modulus $Y=1.131$ TPa and the TOES $D=-2.360$ TPa.
\begin{figure}
  \begin{center}
    \scalebox{1.2}[1.2]{\includegraphics[width=7cm]{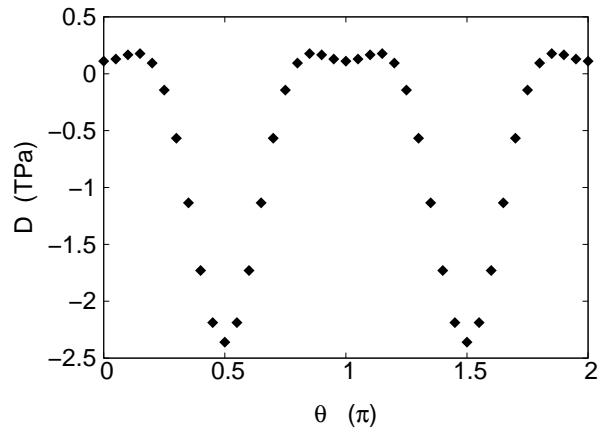}}
  \end{center}
  \caption{The third-order stiffness $D$ in the graphene.}
  \label{fig_D}
\end{figure}

Fig.~\ref{fig_Young} is the dependence of the Young's modulus on the direction angle $\theta$. The average value 
over $\theta$ is about 0.83 TPa. We assume that the experimental measured Young's modulus value $1.0\pm 0.1$ TPa is an average over $\theta$. This assumption can be qualitatively understood in terms that the atomic force microscope tip touches the center of the circular graphene sample; thus generating strains equally in all radial directions. Our calculation is in good agreement with this experimental result. As shown in Fig.~\ref{fig_D}, the average value for TOES $D$ is about -1.3 TPa, which is comparable with the corresponding experimental value -2 $\pm$0.4 TPa. Our result is a little smaller due to the simplicity of our model, which may underestimate the non-linear interaction in the system. The main reason for the underestimation is that the interaction $V_{l}$ can dominate the total interaction in graphene. However, there are some other weaker interactions in the graphene, such as twisting interaction. As these weaker interactions are neglected in present model, the $V_{l}$ with force constant $k_{l}$ underestimates the linear interaction. Consequently, the non-linear force constant from $k_{nl}=3k_{l}$ will underestimate the non-linear interaction in the system.

In conclusion, we calculate the Young's modulus and TOES of the graphene with a simplified system and very simple interaction without any variable parameters and explain the recent experimental results nicely.

We further remark that a large piece of graphene can be regarded as a thin plate. In a plate, the non-linear effect of the third order arises from terms in the elastic energy which are cubic in the strains.\cite{Landau} This non-linear effect leads to a non-linear equation of motion for the system, which indicates the coupling between different phonon modes. In the graphene, the non-linear interaction has exhibited itself in different phenomena. As a direct result of the non-linear interaction, the Raman G mode shows a red-shift with the increase of temperature.\cite{Calizo} Very recently, the non-linear interaction in carbon nanotubes is confirmed to be the origin of the intrinsic localized mode, which leads to the Stone-Wales defect under axial tension.\cite{Shimada} This effect should exist in the graphene.

\textbf{Acknowledgements} The work is supported by a Faculty Research Grant of R-144-000-257-112 of NUS, and Grant R-144-000-203-112 from Ministry of Education of Republic of Singapore, and Grant R-144-000-222-646 from NUS.

\end{document}